\documentclass[leqno]{article}

\usepackage{a4wide}
\usepackage[]{amsfonts} \usepackage[]{amsmath}
\usepackage[]{amssymb} \usepackage[]{latexsym}
\usepackage{graphicx,color} \usepackage{amsthm}
\usepackage{url}

\begin{document}

\theoremstyle{plain}
\newtheorem{theorem}{Theorem}[section] \newtheorem*{theorem*}{Theorem}
\newtheorem{proposition}[theorem]{Proposition} \newtheorem*{proposition*}{Proposition}
\newtheorem{lemma}[theorem]{Lemma} \newtheorem*{lemma*}{Lemma}
\newtheorem{corollary}[theorem]{Corollary} \newtheorem*{corollary*}{Corollary}

\theoremstyle{definition}
\newtheorem{definition}[theorem]{Definition} \newtheorem*{definition*}{Definition}
\newtheorem{example}[theorem]{Example} \newtheorem*{example*}{Example}
\newtheorem{remark}[theorem]{Remark} \newtheorem*{remark*}{Remark}
\newtheorem{hypotheses}[theorem]{Hypotheses} \newtheorem{assumption}[theorem]{Assumption}
\newtheorem{notation}[theorem]{Notation} \newtheorem*{question}{Question}

\newcommand{\ds}{\displaystyle} \newcommand{\nl}{\newline}
\newcommand{\eps}{\varepsilon}
\newcommand{\bE}{\mathbb{E}}
\newcommand{\cB}{\mathcal{B}}
\newcommand{\cF}{\mathcal{F}}
\newcommand{\cA}{\mathcal{A}}
\newcommand{\cM}{\mathcal{M}}
\newcommand{\cD}{\mathcal{D}}
\newcommand{\cH}{\mathcal{H}}
\newcommand{\cN}{\mathcal{N}}
\newcommand{\cL}{\mathcal{L}}
\newcommand{\cLN}{\mathcal{LN}}
\newcommand{\bP}{\mathbb{P}}
\newcommand{\bQ}{\mathbb{Q}}
\newcommand{\bN}{\mathbb{N}}
\newcommand{\bR}{\mathbb{R}}
\newcommand{\barsigma}{\overline{\sigma}}
\newcommand{\VIX}{\mbox{VIX}}
\newcommand{\erf}{\mbox{erf}}
\newcommand{\LMMR}{\mbox{LMMR}}
\newcommand{\cLcir}{\mathcal{L}_{\!_{C\!I\!R}}}
\newcommand{\rhor}{\raisebox{1.5pt}{$\rho$}}
\newcommand{\varphir}{\raisebox{1.5pt}{$\varphi$}}
\newcommand{\taur}{\raisebox{1pt}{$\tau$}}
\newcommand{\spx}{S\&P 500 }

\title{Pricing Path-Dependent Derivatives under Multiscale Stochastic Volatility Models: a Malliavin Representation}

\author{Yuri F. Saporito}

\maketitle

\begin{abstract}
In this paper we derive a efficient Monte Carlo approximation for the price of path-dependent derivatives under the multiscale stochastic volatility models of \cite{fouque_book}. Using the formulation of this pricing problem under the functional It\^o calculus framework and making use of Greek formulas from Malliavin calculus, we derive a representation for the first-order approximation of the price of path-dependent derivatives in the form $\mathbb{E}[\mbox{payoff} \times \mbox{weight}]$. The weight is known in closed form and depends only on the group market parameters arising from the calibration of the multiscale stochastic volatility to the market's implied volatility. Moreover, only simulations of the Black--Scholes model is required. We exemplify the method for a couple path-dependent derivatives.
\end{abstract}

\section{Introduction}

The \textit{multiscale stochastic volatility models} of J.-P. Fouque, G. Papanicolaou, R. Sircar, and K. S{\o}lna are a powerful approach to reconcile stochastic volatility models and computational tractability of pricing and calibration; see, for instance, \cite{fouque_book}. 

Currently, there are two main approaches to deal with the pricing of exotic derivatives in this setting. If the option is path-independent (meaning the payoff depends only on the final value of the spot price) and closed-form solution is available for the zero-order price (which is the price of the option under Black--Scholes model), then the first-order correction is readily available by direct computation of certain Greeks of the zero-order price. In the case of path-dependent payoffs, one would have to rely on numerical solution of partial differential equations (PDEs). This could be done efficiently, but there are issues nonetheless. The PDE is usually in two or three dimensions and the algorithm needs to be run for each single option. Moreover, in order to consider the slow time-scale factor, it is necessary to numerically approximate the Vega and the Vanna (Delta of the Vega) of the zero-order price.

In this paper, we develop a Monte Carlo method to compute the first-order approximation of any\footnote{As we will see, the method here will deal with any discretely-monitored option, which encapsulate all realistic option payoff.} path-dependent option. Moreover, the method requires only simulation of the Black--Scholes model and computation of certain weights related to the Greeks involved in the multiscale correction. Furthermore, once these quantities are simulated, one might compute the first-order approximation for any contract on the same asset.

The paper is organized as follows. The next section introduces the model and the first-order approximation under the path-dependent setting. In Section \ref{sec:path-indep} we derive the Monte Carlo approach for path-independent payoff in order to introduce the main idea, which is developed in detail in Section \ref{sec:main}. Finally, we exemplify the method in Section \ref{sec:numerical}.

\section{Multiscale Stochastic Volatility Models}

The model, under a risk-neutral measure, can be written as

\begin{align}\label{eq:sde_risk_neutral_intro}
\left\{
\begin{array}{l}
dS_t = r S_t dt + f(Y^{\eps}_t,Z^{\delta}_t) S_t dW_t,\\ \\
\ds dY^{\eps}_t = \left(\frac{1}{\eps} \alpha(Y_t^{\eps}) - \frac{1}{\sqrt{\eps}} \beta(Y_t^{\eps}) \Gamma_1(Y^{\eps}_t,Z^{\delta}_t) \right)dt + \frac{1}{\sqrt{\eps}} \beta(Y_t^{\eps}) dW_t^Y, \\ \\
dz^{\delta}_t = \left(\delta c(Z_t^{\delta}) - \sqrt{\delta} g(Z_t^{\delta}) \Gamma_2(Y^{\eps}_t,Z^{\delta}_t) \right)dt + \sqrt{\delta} g(Z^{\delta}_t) dW_t^Z,
\end{array}
\right.
\end{align}
where $(W_t,W_t^Y,W_t^Z)$ is a correlated $\bQ$-Brownian motion. Here $S_t$ denotes the price of the underlying asset at time $t$ and $Y^\eps$ and $Z^\delta$ model, respectively, the fast and slow mean-reverting factors of the asset's volatility.
The functions $\Gamma_1$ and $\Gamma_2$ together completely define the market price of volatility risk and uniquely determine the risk-neutral measure $\bQ$. The usual assumptions are required for the correlations and the functions $f, \alpha, \beta, c, q, \Gamma_1$ and $\Gamma_2$ as shown in \cite{fouque_book}. If $T$ is the typical maturity of options contracts in this market, both $\eps$ and $\delta$ should be thought as small parameters in the sense that $\eps \ll T \ll 1/\delta$.

The path-dependent derivatives we will consider here are of the form $h(x_{t_1}, \ldots, x_{t_n})$ where $0 < t_1 < t_2 < \cdots < t_n = T$. Realistically, these are the most general payoffs available in financial markets. These payoffs are called discretely-monitored, see \cite{fito_saporito_greeks}.

For the sake of brevity, we forward the reader to \cite{fito_multiscale} for an introduction to functional It\^o calculus and for its application to pricing under multiscale stochastic volatility. Moreover, we will \textit{not} use the notation from functional It\^o calculus giving preference to the standard stochastic calculus notation.

From the results presented \cite{fito_multiscale}, one may straightforwardly apply them to the discretely-monitored payoffs and conclude that, for $t \in [t_i, t_{i+1})$, the first-order correction is given by
\small
\begin{align}
P_0(t, x_1, \ldots, x_i, x) &= \bE_{t,x}[e^{-r(T-t)} h(x_1, \ldots, x_i, X_{t_{i+1}}, \ldots, X_{t_n})], \label{eq:p0} \\
P_1^{\eps,\delta}(t, x_1, \ldots, x_i, x)&= \int_t^{t_{i+1}} e^{-r(s-t)} \bE_{t,x}\left[\cH P_0(s, x_1, \ldots, x_i, X_s) \right]du \label{eq:p1}\\
&+ \sum_{j=i+1}^{n-1} \int_{t_j}^{t_{j+1}} e^{-r(s-t)} \bE_{t,x}\left[\cH P_0(s, x_1, \ldots, x_i, X_{t_{i+1}},\ldots, X_{t_{j-1}}, X_s) \right]du, \nonumber
\end{align}
\normalsize
where the operator $\cH$ is
\begin{align}\label{eq:cH}
\cH = 2 V_0^\delta \frac{\partial}{\partial \sigma} + 2 V_1^\delta D_1 \frac{\partial}{\partial \sigma} + V_3^\eps D_1 D_2 \mbox{ and } D_k = x^k \frac{\partial^k}{\partial x^k},
\end{align}
with the derivative $D_k$ always with respect to the last variable of the function. See also Chapter 9 of \cite{fouque_book}. We are suppressing the dependence on the slow factor $z$ for simpler notation. Here $V_0^\delta, V_1^\delta$ and $V_3^\eps$ are small parameters arising from the multiscale model. Moreover, $\bE_{t,x}$ is the expectation conditional to $X_t = x$ and $X$ follows the geometric Brownian motion with volatility $\barsigma$:
$$dX_s = rX_s ds + \barsigma X_s dW_s,$$
where $\barsigma$ is the effective volatility (the average under the stationary distribution of $Y^\eps$).

The goal of this work is to derive a weight $\pi^{\eps,\delta}(t,T)$ that depends only on the model (and not on the payoff $h$) such that the first-order approximation maybe written as
\begin{align}\label{eq:main_formula}
(P_0 + P_1^{\eps,\delta})(t, x_1, \ldots, x_i, x) = \bE_{t,x}[e^{-r(T-t)} h(x_1, \ldots, x_i, X_{t_{i+1}}, \ldots, X_{t_n}) \times \pi^{\eps,\delta}(t,T)].
\end{align}
As we will show, the weight $\pi^{\eps,\delta}(t,T)$ will depend only on the market group parameters $V_0^\delta, V_1^\delta$,  $V_3^\eps$, $\barsigma$ and the Brownian motion $W$. Therefore, in order to estimate the expectation in (\ref{eq:main_formula}) by Monte Carlo, one needs to simulate only $W$ and $X$. It is important to notice that once $(X_{t_{i+1}}, \ldots, X_{t_n})$ is simulated from the Black--Scholes model starting at $X_t=x$, with volatility $\barsigma$ and interest rate $r$, and $\pi^{\eps,\delta}(t,T)$ is computed using the market group parameters and the Brownian motion used in the simulation of $X$, we could use these simulated values to compute the first-order correction of any discretely-monitored (on a subset of $\{t_{i+1},\ldots,t_n\}$), path-dependent derivative.

\section{Derivation for Path-Independent Payoff}\label{sec:path-indep}

In order to introduce the main idea of the article let us consider the simpler case of pricing a derivative with path-independent payoff $h(S_T)$. As it was shown in \cite{fouque_book}, $P^{\eps,\delta}_1$ can be written in closed-form depending only on the Greeks of $P_0$ that compose the operator $\cH$. Since our goal is to study more general payoffs that do not allow for such representations, we will not consider these formulas here. We will take a more general approach that applies even to path-dependent derivatives as we will see in Section \ref{sec:main}.

By Feynman-Kac's Formula, we may write
\begin{align*}
P_1^{\eps,\delta}(t,x) &= \bE_{t,x}\left[\int_t^T e^{-r(s-t)} \cH P_0(s,X_s)ds\right] = \int_t^T e^{-r(s-t)} \bE_{t,x}\left[\cH P_0(s,X_s) \right]ds.
\end{align*}
Using the weight representation for Greeks that we will specify soon, we will be able write:
$$\cH P_0(s,y) = e^{-r(T-s)} \bE_{s,y}\left[h(X_T) \pi_\cH^{\eps, \delta}(s,T)\right],$$
where $\pi_\cH^{\eps, \delta}(s,T)$ will be known in closed form.
Putting these formulas together, we find
\begin{align*}
P_1^{\eps,\delta}(t,x) &= \bE_{t,x}\left[\int_t^T e^{-r(s-t)} \cH P_0(s,X_s)ds\right] = \int_t^T e^{-r(s-t)} \bE_{t,x}\left[\cH P_0(s,X_s)\right]ds\\
&= \int_t^T e^{-r(s-t)} \bE_{t,x}\left[e^{-r(T-s)} \bE_{s,X_s}\left[h(X_T) \pi_\cH^{\eps, \delta}(s,T)\right] \right]ds\\
&= \int_t^T e^{-r(T-t)} \bE_{t,x}\left[ h(X_T) \pi_\cH^{\eps, \delta}(s,T)\right]ds = e^{-r(T-t)} \bE_{t,x}\left[h(X_T)  \int_t^T \pi_\cH^{\eps, \delta}(s,T)ds \right].
\end{align*}
In other words, the first-order approximation can be written as
$$(P_0 + P_1^{\eps,\delta})(t,x) = e^{-r(T-t)} \bE_{t,x}\left[h(X_T) \left(1 +  \int_t^T \pi_\cH^{\eps, \delta}(s,T)ds\right) \right].$$
Therefore, the weight 
$$\pi^{\eps,\delta}(t,T) = 1 +  \int_t^T \pi_\cH^{\eps, \delta}(s,T)ds$$
satisfies Equation (\ref{eq:main_formula}).

\subsection{Derivation of the Weight $\pi^{\eps,\delta}(t,T)$}

In this section we will compute $\pi^{\eps,\delta}(t,T)$ in closed form depending only on the model parameters and the Brownian motion $W$. Remember that the underlying model $X$ follows the Black--Scholes dynamics. Hence, we have available from \cite{FournieMalliavin}, the following formulas
\begin{align*}
D_2 P_0(t,x) &=  e^{-r(T-t)} \bE_{t,x}\left[h(X_T)  \frac{1}{\sigma (T-t)}\left(\frac{(W_T - W_t)^2}{\sigma (T-t)} - \frac{1}{\sigma} - (W_T - W_t)\right) \right],\\
\frac{\partial P_0}{\partial \sigma} (t,x) &=  e^{-r(T-t)} \bE_{t,x}\left[h(X_T) \left(\frac{(W_T - W_t)^2}{\sigma (T-t)} - \frac{1}{\sigma} - (W_T - W_t)\right)\right],\\
\end{align*}
Hence, we define
\begin{align*}
\pi_\Gamma(t,T) &=  \frac{1}{\sigma (T-t)}\left(\frac{(W_T - W_t)^2}{\sigma (T-t)} - \frac{1}{\sigma} - (W_T - W_t)\right),\\
\pi_\sigma(t,T) & = \sigma (T-t) \pi_\Gamma(t,T),
\end{align*}
%we may write
%\begin{align*}
%D_2 P_0(t,x) &=  e^{-r(T-t)} \bE_{t,x}\left[h(X_T) \pi_\Gamma(t,T)  \right],\\
%\frac{\partial P_0}{\partial \sigma} (t,x) &=  e^{-r(T-t)} \bE_{t,x}\left[h(X_T) \pi_\sigma(t,T) \right].
%\end{align*}
An additional application of the same argument used to derive the expressions above yields
\begin{align*}
D_1 \frac{\partial P_0}{\partial \sigma} (t,x) &=  e^{-r(T-t)} \bE_{t,x}\left[h(X_T) \left(\pi_\Delta(t,T)\pi_\sigma(t,T) - \frac{2}{\sigma}\pi_\Delta(t,T) + \frac{1}{\sigma}\right) \right],\\
D_1 D_2 P_0(t,x) &=  e^{-r(T-t)} \bE_{t,x}\left[h(X_T) \frac{1}{\sigma (T-t)}\left(\pi_\Delta(t,T)\pi_\sigma(t,T) - \frac{2}{\sigma}\pi_\Delta(t,T)+ \frac{1}{\sigma}\right) \right],
\end{align*}
where $\pi_\Delta(t,T) = \tfrac{W_T - W_t}{\sigma (T-t)}$. Therefore, if we define
$$\pi_{\Delta \sigma}(t,T) = \pi_\Delta(t,T)\pi_\sigma(t,T) - \frac{2}{\sigma}\pi_\Delta(t,T) + \frac{1}{\sigma},$$
we conclude
$$ \pi_\cH^{\eps, \delta}(t,T) = 2V_0^\delta \pi_\sigma(t,T) + \left(2V_1^\delta + V_3^\eps \frac{1}{\sigma (T-t)}\right) \pi_{\Delta \sigma}(t,T).$$

\begin{remark}
Formula (\ref{eq:main_formula}) might be useful even in this simpler setting when $P_0$ is not know in closed-form and $\cH P_0$ cannot be computed analytically.
\end{remark}

\section{Derivation for Path-Dependent Payoff}\label{sec:main}

In this section we will derive the main result of our paper. We will adapt the argument shown in Section \ref{sec:path-indep} for the case of discretely-monitored, path-independent payoffs. We will follow the same structure of the section above where we first assume the existence of the weight for the operator $\cH$ to derive the weight for the first-order approximation. We will then derive this weight of $\cH$ based on the Malliavin weights of simpler Greeks.

As we did before, assume for now there exists a weight $\pi_{\cH,j}^{\eps, \delta}$ such that, for any $s \in [0,T)$,
$$\cH P_0(s, y_1, \ldots, y_j, y) = e^{-r(T-s)} \bE_{s,y}\left[h(y_1, \ldots, y_j, X_{t_{j+1}}, \ldots, X_{t_n}) \pi_{\cH,j}^{\eps, \delta}(s,T) \right],$$
where $s \in [t_j, t_{j+1})$. Then, from Equation (\ref{eq:p1}),
\small
\begin{align*}
&P_1^{\eps,\delta}(t, x_1, \ldots, x_i, x)= \int_t^{t_{i+1}} e^{-r(s-t)} \bE_{t,x}\left[e^{-r(T-s)} \bE_{s,X_s}\left[h(x_1, \ldots, x_i, X_{t_{i+1}}, \ldots, X_{t_n}) \pi_{\cH,i}^{\eps, \delta}(s,T) \right] \right]ds \\
&+ \sum_{j=i+1}^{n-1} \int_{t_j}^{t_{j+1}} e^{-r(s-t)} \bE_{t,x}\left[e^{-r(T-s)} \bE_{s,X_s}\left[h(x_1, \ldots, x_i, X_{t_{i+1}}, \ldots, X_{t_n}) \pi_{\cH,j}^{\eps, \delta}(s,T) \right] \right]ds\\
&= \int_t^{t_{i+1}} e^{-r(T-t)} \bE_{t,x}\left[h(x_1, \ldots, x_i, X_{t_{i+1}}, \ldots, X_{t_n}) \pi_{\cH,i}^{\eps, \delta}(s,T) \right]ds \\
&+ \sum_{j=i+1}^{n-1} \int_{t_j}^{t_{j+1}} e^{-r(T-t)} \bE_{t,x}\left[h(x_1, \ldots, x_i, X_{t_{i+1}}, \ldots, X_{t_n}) \pi_{\cH,j}^{\eps, \delta}(s,T) \right]ds\\
&=e^{-r(T-t)} \bE_{t,x}\left[h(x_1, \ldots, x_i, X_{t_{i+1}}, \ldots, X_{t_n}) \left(\int_t^{t_{i+1}} \pi_{\cH,i}^{\eps, \delta}(s,T)ds +  \sum_{j=i+1}^{n-1} \int_{t_j}^{t_{j+1}} \pi_{\cH,j}^{\eps, \delta}(s,T)ds  \right)\right].
\end{align*}
\normalsize
Hence,
\small
\begin{align*}
&(P_0 + P_1^{\eps,\delta})(t, x_1, \ldots, x_i, x) \\
&= e^{-r(T-t)} \bE_{t,x}\left[ h(x_1, \ldots, x_i, X_{t_{i+1}}, \ldots, X_{t_n}) \left(1 +  \int_t^{t_{i+1}} \pi_{\cH,i}^{\eps, \delta}(s,T)ds +  \sum_{j=i+1}^{n-1} \int_{t_j}^{t_{j+1}} \pi_{\cH,j}^{\eps, \delta}(s,T)ds\right) \right].
\end{align*}
\normalsize
Therefore, we find that the weight 
$$\pi^{\eps,\delta}(t,T) = 1 +  \int_t^{t_{i+1}} \pi_{\cH,i}^{\eps, \delta}(s,T)ds +  \sum_{j=i+1}^{n-1} \int_{t_j}^{t_{j+1}} \pi_{\cH,j}^{\eps, \delta}(s,T)ds$$
satisfies Equation (\ref{eq:main_formula}).

%\begin{align*}
%P_1^{\eps,\delta}(t, x_1, \ldots, x_i, x) &= \sum_{j=i+1}^n \int_{\tau_{j-1}}^{\tau_j} e^{-r(u-t)} \bE\left[\cH P_0(u, x_1, \ldots, x_i, X_{t_{i+1}},\ldots, X_{t_{j-1}}, X_u) \ | \ X_t=x \right]du\\
%&= \sum_{j=i+1}^n \int_{t_{j-1}}^{t_j} e^{-r(T-t)} \bE\left[\bE\left[h(x_1, \ldots, x_i, X_{t_{i+1}}, \ldots, X_{t_n}) \pi_\cH^{\eps, \delta}(s,T) \ | \ X_s\right]\ | \ X_t=x \right]ds\\
%&= e^{-r(T-t)} \bE\left[h(x_1, \ldots, x_i, X_{t_{i+1}}, \ldots, X_{t_n}) \sum_{j=i+1}^n \int_{t_{j-1}}^{t_j} \pi_\cH^{\eps, \delta}(s,T) ds \ | \ X_t=x \right]\\
%&= e^{-r(T-t)} \bE\left[h(x_1, \ldots, x_i, X_{t_{i+1}}, \ldots, X_{t_n}) \int_t^T \pi_\cH^{\eps, \delta}(s,T) ds \ | \ X_t=x \right].
%\end{align*}

\subsection{Derivation of the Weight $\pi^{\eps,\delta}(t,T)$}\label{sec:weight_path_dependent}

A straightforward application of the results in \cite{FournieMalliavin} and \cite{mark_davis_malliavin} gives us, for each $j \in \{1,\ldots,n-1\}$ and $s \in [t_j,t_{j+1})$,
\begin{align*}
D_1D_2 P_0(s, y_1, \ldots, y_j, y) &= \bE_{s,y}\left[e^{-r(T-t)} h(y_1, \ldots, y_j, X_{t_{j+1}}, \ldots, X_{t_n}) \pi_{3,j}(s,T) \right],\\
\frac{\partial}{\partial \sigma} P_0(s, y_1, \ldots, y_j, y) &= \bE_{s,y}\left[e^{-r(T-t)} h(y_1, \ldots, y_j, X_{t_{j+1}}, \ldots, X_{t_n}) \pi_{\sigma,j}(s,T)\right],\\
D_1\frac{\partial}{\partial \sigma} P_0(s, y_1, \ldots, y_j, y) &= \bE_{s,y}\left[e^{-r(T-t)} h(y_1, \ldots, y_j, X_{t_{j+1}}, \ldots, X_{t_n}) \pi_{\Delta\sigma,j}(s,T) \right],
\end{align*}
where
\begin{align*}
\pi_{3,j}(s,T) &= \pi_{\Delta,j}^3(s,T) - \pi_{\Delta,j}^2(s,T) - \frac{1}{\sigma^2}(3\pi_{\Delta,j}(s,T) - 1) \int_s^T a^2(t,u)du, \\
\pi_{\Delta,j}(s,T) &= \frac{1}{\sigma}\int_s^T a(s,u) dW_u,\\
\pi_{\sigma,j}(s,T) &= \frac{1}{\sigma}\left( - \int_s^T b(s,u) du + F_j \int_s^{t_{j+1}} b(s,u) dW_u + \sum_{k=j+1}^{n-1} F_k \int_{t_k}^{t_{k+1}} b(s,u) dW_u\right) ,\\
\pi_{\Delta\sigma,j}(s,T) &= \pi_{\sigma,j}(s,T)\pi_{\Delta,j}(s,T) \\
& \hspace*{-30pt}- \frac{1}{\sigma^2} \left(\int_s^T b(s,u) dW_u + F_j \int_s^{t_{j+1}} a(s,u) b(s,u) du + \sum_{k=j+1}^{n-1} F_k \int_{t_k}^{t_{k+1}} a(s,u) b(s,u) du \right) ,\\
F_j &= -\sigma(t_{j+1} - s) + W_{t_{j+1}} - W_s, \quad F_k = -\sigma(t_{k+1} - t_k) + W_{t_{k+1}} - W_{t_k},
\end{align*}
with
\begin{align*}
\int_s^{t_k} a(s,u) du = 1, \quad \int_s^{t_{j+1}} b(s,u) du = 1 \mbox{ and } \int_{t_k}^{t_{k+1}} b(s,u) du = 1,
\end{align*}
for all $k \in \{j+1,\ldots,n-1\}$. Finally, we conclude that
$$ \pi_{\cH,j}^{\eps, \delta}(s,T) = 2V_0^\delta \pi_{\sigma,j}(s,T) + 2V_1^\delta \pi_{\Delta\sigma,j}(s,T) + V_3^\eps \pi_{3,j}(s,T).$$

\begin{remark}\label{rmk:future}
We could also consider different payoff types following the different (yet similar) weight formulas of \cite{fito_saporito_greeks} (weakly path-dependent), \cite{FournieMalliavin} (continuous Asian option)
 and \cite{Gobet2003} (barrier and lookack options).
\end{remark}

\begin{remark}\label{rmk:choice_ab}
A simple choice for $a$ and $b$ for each fixed $s \in [t_j,t_{j+1})$ is
$$a(s,u) = \frac{1}{t_{j+1} - s} 1_{[s,t_{j+1}]}(u) \mbox{ and } b(s,u) = \frac{1}{t_{j+1} - s}1_{[s, t_{j+1})}(u) + \sum_{k=j+1}^{n-1} \frac{1}{t_{k+1} - t_k}1_{[t_k, t_{k+1})}(u).$$
Other possibilities could be considered, \cite{FournieMalliavin2}.
\end{remark}

\subsection{Numerical Implementation}\label{sec:numerical_imp}

It is very common that $t_j = j \Delta t$, for some small $\Delta t$ (usually one day). In this case and taking $t = 0 = t_0$, we consider the following approximation
\begin{align*}
\pi^{\eps,\delta}(0,T) = 1 +  \sum_{j=0}^{n-1} \int_{t_j}^{t_{j+1}} \pi_{\cH,j}^{\eps, \delta}(s,T)ds \approx  1 +  \sum_{j=0}^{n-1} \pi_{\cH,j}^{\eps, \delta}(t_j,T) \Delta t.
\end{align*}
Using the choices of $a$ and $b$ from Remark \ref{rmk:choice_ab}, the weights can be written as
\begin{align*}
\pi_{\Delta,j}(t_j,T) &= \frac{1}{\sigma \Delta t} \Delta W_{t_{j+1}},\\
\pi_{3,j}(t_j,T) &= \pi_{\Delta,j}^3(t_j,T) - \pi_{\Delta,j}^2(t_j,T) - \frac{1}{\sigma^2\Delta t}\left(3\pi_{\Delta,j}(t_j,T) - 1\right),\\
\pi_{\sigma,j}(t_j,T) &=  \frac{1}{\sigma} \left(- (n-j)+ \frac{1}{\Delta t}\sum_{k=j}^{n-1} F_k \Delta W_{t_{k+1}}\right) ,\\
\pi_{\Delta\sigma,j}(t_j,T) &= \pi_{\sigma,j}(t_j,T)\pi_{\Delta,j}(t_j,T) - \frac{1}{\sigma^2 \Delta t} \left((W_T - W_{t_j}) +  F_j \right) ,\\
F_k &= -\sigma\Delta t + \Delta W_{t_{k+1}},
\end{align*}
where $\Delta W_{t_{k+1}} = W_{t_{k+1}} - W_{t_k}$. 
%Therefore, one can simplify even further and find
%\begin{align*}
%&\pi^{\eps,\delta}(0,T) \approx 1 + 2V_0^\delta \left(-\frac{T}{2}(n-1) - \sigma T W_T + \sigma \Delta t \sum_{j=0}^{n-1} W_{t_j} + \sum_{j=0}^{n-1} \sum_{k=j}^{n-1} \Delta W_{t_{k+1}}^2\right)\\
%&+ 2V_1^\delta \left(T - \frac{1}{\sigma} W_T + \sum_{j=0}^{n-1} \frac{1}{\sigma}(j - n) \Delta W_{t_{j+1}} - \Delta W_{t_{j+1}} (W_T - W_{t_j}) - \frac{1}{\sigma}(W_T - W_j) + \frac{1}{\sigma \Delta t} \Delta W_{t_{j+1}}  \sum_{k=j}^{n-1} \Delta W_{t_{k+1}}^2 \right)\\
%&+ V_3^\eps \left(\frac{1}{\sigma^3 \Delta t^2} \sum_{j=0}^{n-1} \Delta W_{t_{k+1}}^3 - \frac{1}{\sigma^2 \Delta t} \sum_{j=0}^{n-1} \Delta W_{t_{k+1}}^2 - \frac{3}{\sigma \Delta t} W_T + n\right)\\
%&= 1 + 2V_0^\delta \left(-\frac{T}{2}(n-1) - \sigma T W_T + \sigma \Delta t \sum_{j=0}^{n-1} W_{t_j} + \sum_{j=0}^{n-1} (j+1) \Delta W_{t_{k+1}}^2\right)\\
%&+ 2V_1^\delta \left(T - \frac{1}{\sigma} W_T + \sum_{j=0}^{n-1} \left\{ \frac{1}{\sigma}(j - n) \Delta W_{t_{j+1}} - \left(\Delta W_{t_{j+1}} + \frac{1}{\sigma} \right) (W_T - W_{t_j}) + \frac{1}{\sigma \Delta t} W_{t_{j+1}}  \Delta W_{t_{j+1}}^2 \right\}\right)\\
%&+ V_3^\eps\left(  - \frac{3}{\sigma \Delta t} W_T + n + \frac{1}{\sigma^2 \Delta t}\sum_{j=0}^{n-1} \left( \frac{1}{\sigma \Delta t} \Delta W_{t_{k+1}}^3 -  \Delta W_{t_{k+1}}^2 \right) \right),
%\end{align*}

\section{Numerical Example}\label{sec:numerical}

\begin{figure}[h!]
\centering
\includegraphics[scale=0.25]{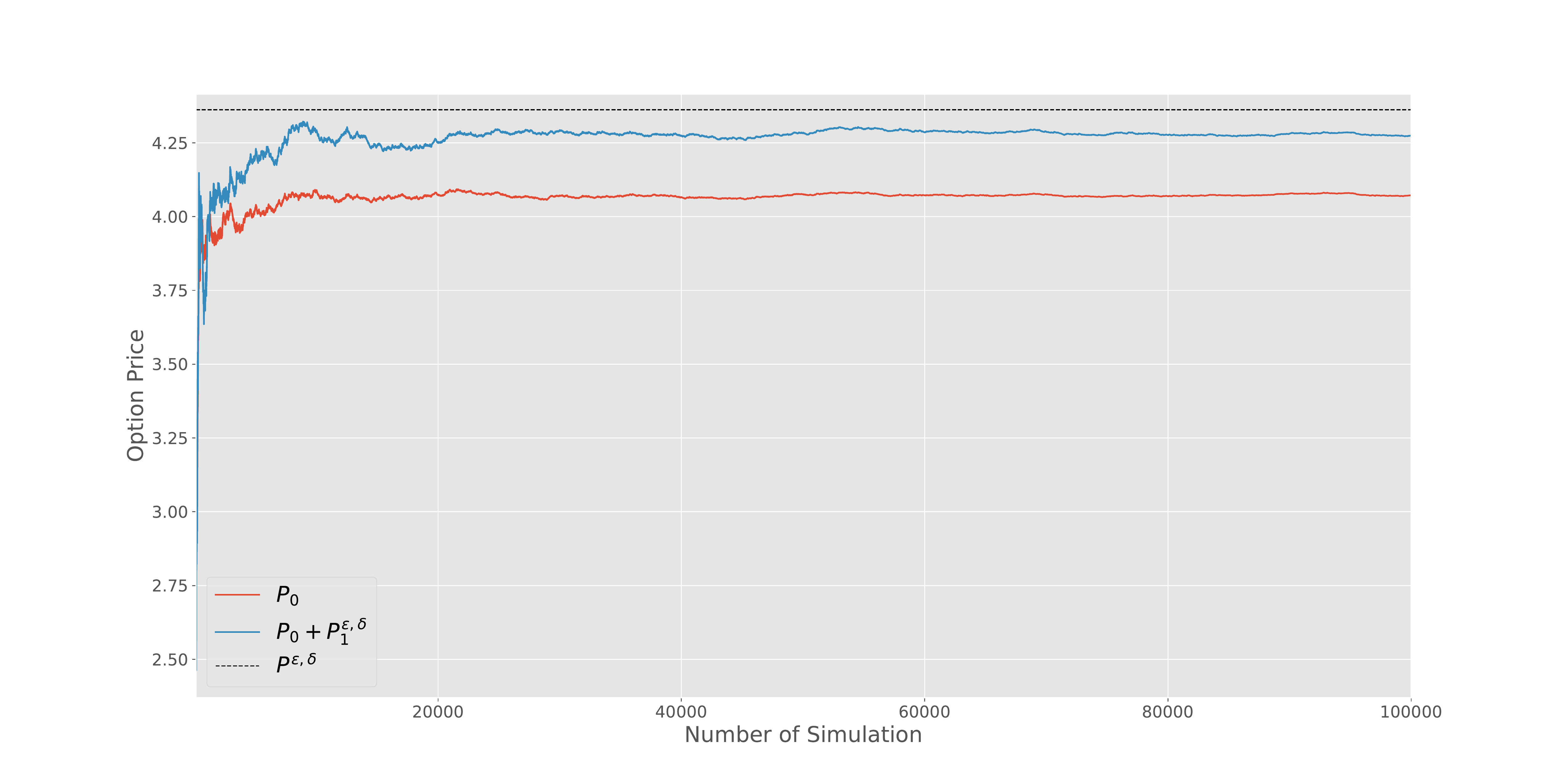}
\caption{Convergence Plot for the Asian option} \label{fig:asian}
\includegraphics[scale=0.25]{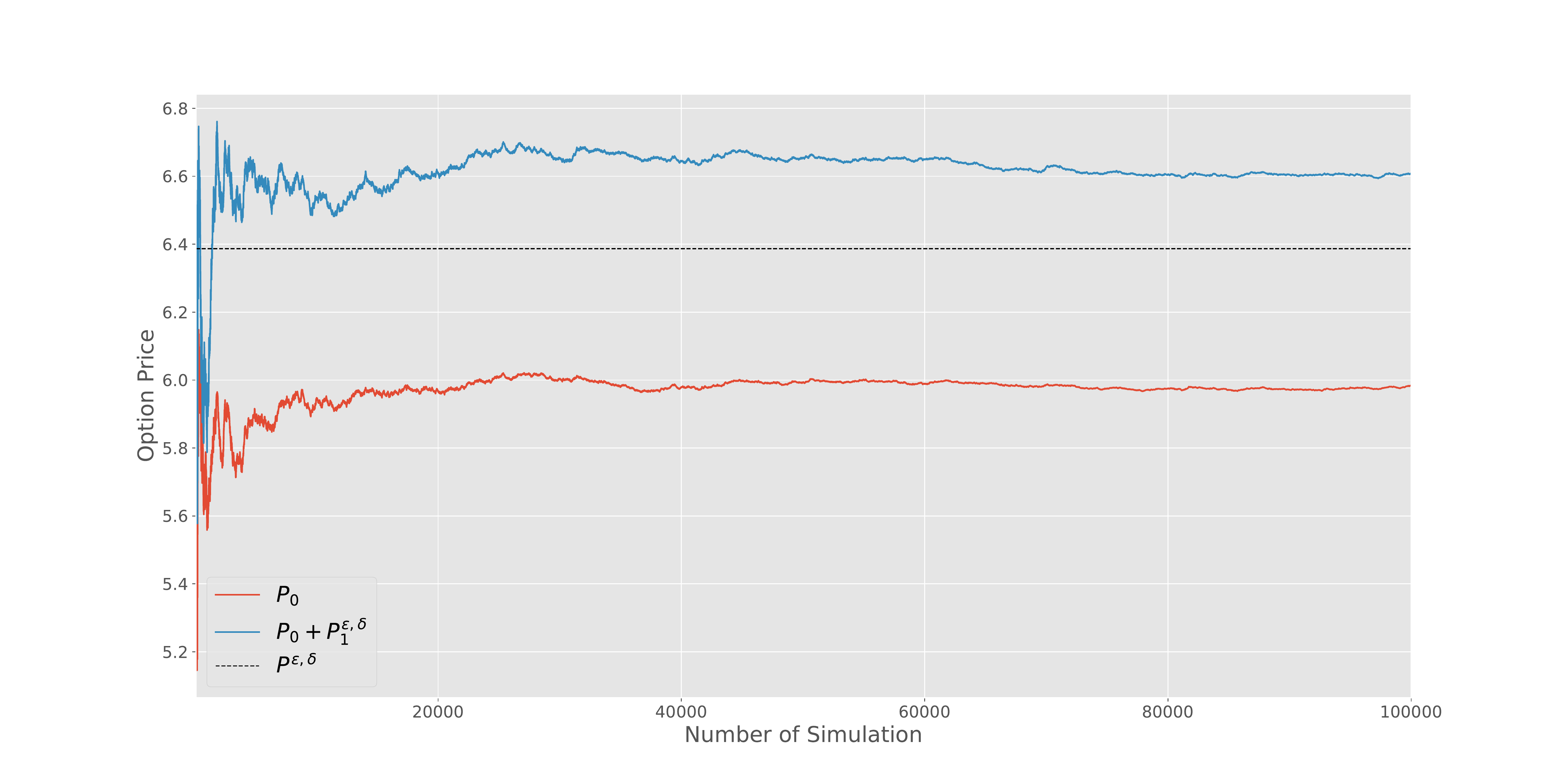}
\caption{Convergence Plot for the Barrier option} \label{fig:barrier}
\end{figure}

In this section, we will exemplify the method developed in this paper. Consider the following discretely-monitored, path-dependent derivatives:
\begin{align*}
&\mbox{Asian Call}: \ h(x_{t_1},\ldots,x_{t_n}) = \left(\frac{1}{n} \sum_{k=1}^n x_{t_i} - K\right)^+,\\
&\mbox{Up-and-Out Call}:  h(x_{t_1},\ldots,x_{t_n}) = \left(x_{t_n} - K\right)^+ 1_{\max\{x_{t_1},\ldots,x_{t_n}\} \leq H},
\end{align*}
where $K$ is the strike, $\{t_i\}$ are the days of monitoring and $T = t_n$ is the maturity. We consider $\alpha(y) = m_1 - y$, $c(z) = m_2 - z$, $\beta(y) = \sqrt{2}\nu_1$, $g(y) = \sqrt{2}\nu_2$, $\Gamma_1 = \Gamma_2 = 0$ and $f(y,z) = \sigma e^{y+z}$. Moreover, we consider the approximation for the weight $\pi^{\eps,\delta}$ shown in Section \ref{sec:numerical_imp}. Furthermore, we denote by $N$ the number of Monte Carlo simulation to estimate the expectation in (\ref{eq:main_formula}). The results are shown in Figure \ref{fig:asian}. We show the convergence plot of the Monte Carlo estimator for zero-order price, the first-order correction and the price under the full model (\ref{eq:sde_risk_neutral_intro}) using $10^6$ simulation paths. We see in both cases the correction is a better approximation of the ``true" price.

Notice that one could apply numerical methods to solve the PDE corresponding to the continuous-time limit of payoff $h$. However, not just this would be a two-dimensional PDE for the Asian option case, but it would require to numerically compute the Vega and $D_1$Vega of the zero-order price, since it is not available in closed-form in both cases.

\begin{table}[h!]
\centering
\small
\begin{tabular}[t]{lr}
Param. & Value \\
\hline
$m_1$ & 0.0\\
$m_2$ & 0.0\\
$\nu_1$ & 0.01\\
$\nu_2$ & 0.01\\
$z$ & 0.0
\end{tabular}
\hspace{0.1cm}
\begin{tabular}[t]{lr}
Param. & Value \\
\hline
$\rho_{SY}$ & -0.5\\
$\rho_{SZ}$ & 0.5\\
$\rho_{YZ}$ & 0.3\\
$\sigma$ & 0.2 \\
$r$ & 0.02
\end{tabular}
\hspace{0.1cm}
\begin{tabular}[t]{lr}
Param. & Value \\
\hline
$\barsigma$ & 0.2020\\
$V_3^\eps$ & $-1.8526 \times 10^{-5}$\\
$V_1^\delta$ & $-4.5397 \times 10^{-5}$\\
$V_0^\delta$ & 0.0\\
$S_0$ & 100
\end{tabular}
\hspace{0.1cm}
\begin{tabular}[t]{lr}
Param. & Value \\
\hline
$T$ & 0.5 \\
$K$ & 100 \\
$H$ & 150 \\
$n$ & 100 \\
$t_i$ & $iT/n$\\
$N$ & $10^5$
\end{tabular}
\caption{Model and Options Parameters.}
\end{table}

\clearpage

\section{Conclusion}

We have developed a representation for the first-order correction of path-dependent option prices under the multiscale stochastic volatility model of \cite{fouque_book}. This representation is of the form $\mathbb{E}[\mbox{payoff} \times \mbox{weight}]$, where the weight depends only on the market group parameters, which are straightforwardly calibrated to the implied volatility surface. This formula could then be used to price any realistic path-dependent exotic option.

The only numerical method available to compute this first-order correction is numerical methods to solve PDEs, but it is not available for the generality of payoffs considered here and it would require to overcome some numerical challenges due to high dimensions and lack of smoothness depending on the pyaoff, see Chapters 6 and 9 of \cite{fouque_book}. The formula developed in this paper does not suffer from any of these drawbacks.

Future research could be carried out in the direction of Remark \ref{rmk:future}. Moreover, one could also consider different auxiliary functions $a$ and $b$ showed in Section \ref{sec:weight_path_dependent}. Furthermore, one could consider the possibility of development similar formula for the second-order correction, see \cite{fouque_second_order}.

 \bibliographystyle{plain}

\end{document}